\documentclass[twocolumn,noshowpacs,preprintnumbers,amsmath,amssymb]{revtex4}
\def\bi{\bibitem}
\def\prl{Phys. Rev. Lett.}\def\pr{Phys. Rev.}

\def\ba{\begin{eqnarray}}
\def\ea{\end{eqnarray}}
\def\la{\langle}\def\ra{\rangle}

\def\be{\begin{eqnarray}}
\def\ee{\end{eqnarray}}

\def\del{\partial}

\def\He#1{{}^{#1}\mbox{He}}
\def\L{{\cal L}}

\usepackage{graphicx}% Include figure files
\usepackage{dcolumn}% Align table columns on decimal point
\usepackage{bm}% bold math

\begin{document}

\preprint{}

\title{QCD and Nuclei}% Force line breaks with \\

\author{Mannque Rho}
\vskip 0.1cm
% \altaffiliation[ Also at ]{School of Physics, Korea
%Institute for Advanced Study,
% Seoul, Korea
% }%Lines break automatically or can be forced with \\
\email{rho@spht.saclay.cea.fr}
\affiliation{%
Service de Physique Th\'eorique, CEA Saclay, 91191 Gif-sur-Yvette,
France
}%
\affiliation{School of Physics, Korea Institute for Advanced
Study, Seoul, Korea }

%\affiliation{


\begin{abstract}
I discuss how the fundamental theory of strong interactions given
by QCD can be exploited to make accurate predictions for the solar
fusion processes involving few-nucleon systems and to provide a
novel description of nuclear matter as well as excitations near
the chiral phase transition. The tool used is effective field
theory combined with the modern technique of renormalization group
flow. The notion of ``BR scaling" is reinterpreted in terms of the
vector manifestation \`a la Harada and Yamawaki. It is argued that
the vector manifestation scenario predicts a QCD phase structure
much richer than the standard scenario discussed up to date in the
literature.

%An article usually includes an abstract, a concise summary of the work
%covered at length in the main body of the article. It is used for
%secondary publications and for information retrieval purposes. Valid
%PACS numbers may be entered using the \verb+\pacs{#1}+ command.
\end{abstract}

\pacs{Valid PACS appear here}% PACS, the Physics and Astronomy
%Use showkeys class option if keyword
\maketitle

\section{\label{intro} Introduction}
There are broadly two ways the fundamental theory of strong
interactions, QCD, can figure importantly in nuclear physics: The
first is to legitimize the place that nuclear physics occupies
within the framework of the Standard Model and the second is to
enable one to make precise predictions for processes involving
nuclei that play a crucial role in certain fundamental processes.
In this talk I will not dwell on the first issue although it has
attracted most of the recent efforts and simply refer to a recent
review~\cite{beaneetal}. I shall instead focus on the second
issue. It turns out that it is in the latter where several Korean
theorists have made significant contributions. I am particularly
proud of their contributions and feel honored to talk about them
at this occasion of the 30th anniversary of Nuclear Physics
Division of the Korean Physical Society.

I shall discuss two issues of fundamental nature here. One is the
precise calculation of the weak processes that take place in the
Sun that are relevant to the solar neutrino problem and the
question of neutrino masses. Here we are dealing with light nuclei
dilute in number density. The other is the property of hadronic
matter under extreme conditions of high temperature and/or high
density. The high temperature property is connected with the Early
Universe and the high density with the interior of compact stars.
QCD should describe what happens under such conditions, some which
are currently accessible to the usual laboratory conditions and
some not. In principle QCD should describe both -- the dilute
system and the dense system -- within the same framework. However
performing such a feat starting from QCD in its elegant form is
impossible at the moment. Hence we are forced to resort to a
certain number of strategies that rely on effective theories, in
particular effective field theories (EFTs). Now in effectuating
EFTs, the two cases I mentioned above require somewhat different
strategies which in principle should be connected but how they are
related is not well understood at the moment.

In a nut-shell, the notion of EFTs that I shall exploit here is
summarized by Weinberg's ``theorem"~\cite{wein-theorem} which in
the present context states that at low energy/momentum, QCD can be
encapsulated in an effective field theory with a suitable set of
colorless fields subject to the symmetries and invariance required
by QCD. This ``theorem" has not yet been proven but it seems
highly unlikely that it cannot be correct. In what follows, I will
assume that doing an EFT at low energy in the spirit of Weinberg's
theorem does indeed amount to doing QCD and discuss how it can be
implemented in confronting Nature.
\section{Effective field theories (EFTs)}
\subsection{Setting up an EFT}
In QCD, an effective field theory is defined below an arbitrary
but suitably chosen scale $\Lambda$ in terms of a set of effective
degrees of freedom that delineate the EFT from a fundamental
theory defined above $\Lambda$ in terms of fundamental degrees of
freedom. Below $\Lambda$ the relevant degrees of freedom are
hadrons with masses less than the cutoff scale and above $\Lambda$
they are the quarks and gluons of QCD. In order for an EFT to
accurately represent QCD, it has to be matched to QCD at the
scale, say, $\Lambda_M$. This can be done by matching physical
observables, say, the current-current correlators, expressed in
terms of the relevant degrees of freedom in both sectors. How this
can be done has been developed and explained in detail by Harada
and Yamawaki in ~\cite{HY:PR} using hidden local symmetry
Lagrangian. One of the important points that one learns from this
development is that when the correlators are matched in medium at
the scale $\Lambda_M$, that is, at a finite temperature and/or
finite density, the parameters of the effective Lagrangian are
parametrically dependent on temperature and/or density as well as
of course on the scale $\Lambda_M$. In what precise form the
dependence takes is not known within the framework of EFTs. They
may be determined either on lattice or by experiments if
available. We note that most of the EFT calculations that start
with an effective Lagrangian defined in the matter-free vacuum do
not have this ``intrinsic dependence" whereas the approach based
on Brown-Rho scaling~\cite{BR91} does. This intrinsic dependence
turns out to play a pivotal role for the arguments given below.
The other important point is that the (Wilsonian) matching with
QCD leads to the result that in the chiral limit (that is, when
quark masses are set to zero) the system must flow to what Harada
and Yamawaki call ``vector manifestation (VM)" fixed
point~\cite{HY:VM,HY:PR} as the critical temperature or density is
approached.  At the VM, the mass of the vector mesons vanishes
with the gauge coupling $g$ and the pion decay constant $f_\pi$
going to zero and the longitudinal components of the vector mesons
$\rho_\parallel$ (i.e., the de-Higgsed scalars) joining, for three
flavors~\footnote{For two flavors, the situation is a bit more
subtle. We will ignore this subtlety here.}, the octet
pseudoscalar Goldstone bosons, $\pi$. The bottom line of this
observation is that at least in the chiral limit, the chiral
restoration phase transition \`a la VM is basically different from
the standard sigma model one. There is no lattice QCD confirmation
of this scenario but it is likely that it will be tested. In this
talk, I will assume that the vector manifestation is a viable
mechanism and discuss what comes out of the picture. It will turn
out that one can make some interesting predictions ranging from
light nuclear systems to dense stellar systems.
\subsection{{\em More effective} effective field theory or MEEFT}
The ultimate goal of an EFT would be to start with an effective
Lagrangian as defined above and derive all properties of few-body
as well as many-body systems in a systematic power counting
scheme, without the necessity to resort to empirical data once a
minimum (initial) set of parameters are fixed. Such a program
would have to deal with not only the basic two-nucleon
interactions but also many-body interactions, accounting for the
formation of the Fermi sea which is in a sense a quantum critical
phenomenon and arriving at a Fermi liquid structure at normal
nuclear matter density. The ultimate objective would then be to
arrive at the chiral phase transition expected in QCD in the
chiral limit. Such an ambitious program does not yet exist and
perhaps makes no sense just as it makes no sense to attempt to
derive crystals starting from QED.

Now even in the framework of EFT, if one were to adhere strictly
to the order-by-order power counting of the EFT series (one might
call this ``purist EFT"), one could not make much progress. All
that has been accomplished up to now is to postdict things on
two-nucleon systems and perhaps a little bit of three-nucleon
systems but doing anything beyond in the mass number, unless one
relaxes the strict rule of power counting, is at present out of
reach. Even for the few-nucleon systems that can be handled, I do
not see much more than just reproducing the phenomena well
understood in the standard nuclear physics approach based on
realistic potentials. It is not clear that this line of efforts
would lead to anything new or even useful in deeper understanding
of complex nuclear dynamics.

We take an alternative approach that takes full advantage of the
tremendous accuracy achieved by the standard nuclear physics
approach (SNPA). The sophisticated SNPA which has been
unquestionably successful in correlating a large number of nuclear
phenomena~\cite{SNPA} is based on ``realistic" nuclear potentials
-- typically two- and three-nucleon interactions -- that are fit
to a large number of scattering as well as spectroscopic data over
a wide range of nuclei. By realistic nuclear potentials, I will
understand those potentials that have the long-range part
constrained by chiral symmetry, that is, pionic properties and the
short-range part determined by experiments. Since
low-energy/momentum processes should not be sensitive to the
detailed structure of the short-range part, the short-distance
part of the interaction could not be uniquely determined by
low-energy data. The issue is how to formulate the EFT such that
transition matrix elements that sample $also$ short-distance
components -- though with much less importance -- can be made
accurate. The way to accomplish this is to incorporate the SNPA
into the framework of effective field theories, while preserving
the power counting of the EFT. This allows one to carry over the
immense success of the SNPA and make the scheme as
model-independent and unambiguous as possible. Note that the SNPA
by itself does not lend itself to a systematic assessment of
errors committed in the calculation. This is because there is no
clear way to establish that what is obtained is not subject to
uncontrolled corrections. The power of an EFT is precisely in
providing the means to assess the error committed in making
approximations and making it possible to systematically correct
the errors. Let me call this version of EFT which combines the
powers of the SNPA and EFT {\it more effective} effective field
theory (MEEFT)~\footnote{I must mention here as a side remark that
this terminology is entirely mine, merely meant to tickle, and not
to ``anger," the aficionados of the ``purist EFT." My coworkers
like to call it EFT*.}. What it means will be clearer later.
\section{EFT predictions for light nuclei}
\subsection{Solar fusion process}
As a case where one can work with an effective Lagrangian
constructed in the vacuum at zero temperature and density matched
to QCD at the chiral scale near $\Lambda_\chi\sim 4\pi f_\pi\sim
1$ GeV, consider the solar fusion processes
 \be pp:&&\ \ \ p+p\rightarrow d + e^+ +\nu_e\,,
\label{pp}
\\
hep:&&\ \ \ p+\He3 \rightarrow \He4 + e^+ + \nu_e\,. \label{hep}
\ee These two processes represent the lowest- and highest-energy
neutrino production processes. The MEEFT enables one to treat
these processes involving different numbers of nucleons on the
same footing~\cite{PMetal2001,PMetal2,PMetallong} and further to
make theoretical error estimates. The reactions (\ref{pp}) and
(\ref{hep}) figure importantly in astrophysics; they have much
bearing upon issues of great current interest such as, for
example, the solar neutrino problem and non-standard physics in
the neutrino sector. Since the thermal energy of the interior of
the Sun is of the order of keV, and since no experimental data is
available for such low-energy regimes, one must rely on theory for
determining the astrophysical $S$-factors of the solar nuclear
processes.

Since the momentum transfer involved is small, these processes are
dominated by the Gamow-Teller matrix element given by the
long-wavelength limit of the spatial component of the axial
current operator. It turns out that the Gamow-Teller operator is
particularly difficult to pin down if its matrix element is
``accidentally" suppressed as in the case of the $hep$ process.
One can understand this in terms of what is known as ``chiral
filter mechanism"~\cite{KDR,chiralfilterMR}. In an effective field
theory in which chiral symmetry plays an essential role as in
nuclei, corrections to the leading single-particle process in the
transition matrix elements are strongly dominated by soft-pion
exchanges whenever they are allowed by symmetry and kinematics. A
consequence is that the subleading corrections remain small
compared to the leading term and can be estimated reasonably well
by going to one order higher: the total amplitude can be computed
with a great deal of accuracy. The space part of the vector
current~\cite{PMR} and the time part of the axial
current~\cite{KDR} belong to this class of operators. However if
the soft-pion exchanges are suppressed by symmetries independently
of kinematic conditions as in the case of the time component of
the vector current and the space component of the axial current,
then leading corrections to the single-particle operators are
sensitive to effects of short-distance nature that are not easily
accessible by low-energy theorems and hence require chiral
non-perturbative treatments. This chiral filter mechanism has been
given a justification in terms of a chiral expansion at low orders
of chiral perturbation~\cite{chiralfilterMR} and also in a
different and more general approach~\cite{serot-walecka}.

It is the lack of the chiral-filter aspect that has made the
calculation of the $hep$ process notoriously difficult for a long
time. The main operator for both (\ref{pp}) and (\ref{hep}) --
which is the Gamow-Teller operator -- is unprotected by the chiral
filter. It turns out however that the leading GT single-particle
matrix for (\ref{pp}) -- which is unsuppressed by symmetry
considerations -- accounts for the bulk of the transition matrix
element, so the short-distance corrections uncontrolled by chiral
symmetry remain in the range of only a few percents at most. This
is not the case for the $hep$ process. Because the leading
single-particle matrix element involved here is highly suppressed
``accidentally" by the symmetry mismatch between the initial and
final wave functions, the corrections play a very important role,
making orders of magnitude uncertainty in the $S$ factor depending
on how they are computed as explained in detail in
~\cite{PMetallong}.

Let me quote the results here for the $pp$ and $hep$ for the sake
of completeness as they are the values to be used by
astrophysicists for solar neutrino issues and explain in the next
subsection for nuclear and hadron physicists how the MEEFT
proposed above can resolve this difficulty and allow the
calculation of the $pp$ process with an unprecedent accuracy and
that of the $hep$ process within $\sim 15$\% accuracy which
represents a giant improvement over the past results~\footnote{I
would like to reiterate my challenge to the aficionados of the
``purist EFT" to come up with a parameter-free prediction for the
$hep$ $S$-factor. The first person who succeeds in doing so within
the coming year will be offered a bottle of a great French wine.}:
 \be
S_{pp}(0) &=& 3.94\!\times\!(1 \pm 0.004) \!\times\!10^{-25}\
\mbox{MeV-b},\label{Spp}\\
S_{hep}(0) &=& (8.6 \pm 1.3)\!\times\! 10^{-20} \
\mbox{keV-b}.\label{Shep}
 \ee
\subsection{The strategy of MEEFT}
Before I go into the details of this subsection, let me do away
with other terms that enter mainly into the $hep$ calculation. In
the weak matrix elements that contribute to both (\ref{pp}) and
(\ref{hep}), the vector matrix elements are without any
uncertainty: They are protected by low-energy theorems. However in
the axial current sector, the time component of the axial current
does make a significant contribution for the $hep$ whereas it is
negligible for the $pp$. In most of the past works with the
exception of the recent SNPA calculation of \cite{marcuccietal},
the axial charge component has not been properly treated. This
component, however, is protected by the chiral filter as mentioned
above, so its calculation can be made fairly reliably. Thus I will
focus on the Gamow-Teller operator which is the only difficult
operator to handle for the processes involved.

To set up the strategy, one first has to pick the relevant degrees
of freedom. Since the nucleons are the essential ingredient of the
nuclear systems, they figure in the theory although the mass scale
involved is of the same scale as the chiral scale $\Lambda_\chi
\sim 1 $ GeV. The pion is the lightest excitation whose mass is
non-zero because of the explicit chiral symmetry breaking due to
the quark masses. Its actual mass is $\sim 140$ MeV. Now if one is
interested in very low-energy and low-momentum processes, say at
most a few MeV, like (\ref{pp}) and (\ref{hep}), then the pion
mass can be considered ``high" and in this case, one might
integrate out the pion as well~\cite{beaneetal}. If the pion is
integrated out with its effect incorporated into higher dimension
terms in the effective Lagrangian, we lose however the power of
the chiral filter consideration and thereby lose certain
predictivity associated with current algebras. We shall therefore
keep the pion in the theory. If the matter density is high, then
as we shall see below, the vector degrees of freedom (such as the
$\rho$ and $\omega$) cannot be ignored since their masses become
small due to the ``vector manifestation" of chiral
symmetry~\cite{HY:PR}. However the systems that figure in
(\ref{pp}) and (\ref{hep}) are dilute and hence the vector mesons
remain massive, so can be integrated out. But this does not mean
that they are not important, {\it particularly if the process is
not protected by the chiral filter}. Nuclear processes generically
experience a wide range of scales as nucleons in strong
interactions sample all distances with varying degrees of
importance and hence when the massive particles are integrated out
their effects transferred to higher-dimension operators cannot be
ignored \`a priori. How those effects are taken into account
differentiate various approximations made in EFT calculations.

The computation of a transition matrix element between the initial
$|i\ra$ and final $|f\ra$ states involves the full account of
interactions between the nucleons and one insertion of an external
field or current. In the Weinberg scheme~\cite{weinberg}, this
amounts to computing the interaction potential and the current
vertex in a series of ``irreducible graphs" and summing to all
orders the ``reducible graphs" with the potential appearing as the
kernel. The power counting goes into the irreducible graphs which
are then computed to a certain order $N$. If one were to compute
to all orders in the irreducible sector, then the procedure would
give an exact answer~\footnote{There could be a problem of
consistency in power counting if this program is executed to a
finite $N$ as discussed in \cite{beaneetal} but if summed to all
orders, this problem would be avoided by fiat.}. One is however
limited to a finite $N$ because of the proliferation of high-order
diagrams and of the number of consequent counter terms that cannot
be fixed (a feature that is characteristic of the so-called
non-renormalizable theory). This limitation is circumvented in
MEEFT by resorting to the potentials that are phenomenologically
fitted to an ensemble of experiments. Such a ``realistic
potential" must have the correct feature at long distance required
by chiral symmetry of QCD, namely one and possibly two-pion
exchanges constrained by low-energy theorems associated with
chiral symmetry. The short-distance components of the variety of
realistic potentials fitted to a set of low energy data, say,
$E\lesssim 300$ MeV, can however differ widely from one potential
to another. It is also in these parts where off-shell ambiguities
could figure strongly. This means that when one probes
short-distance effects with such potentials one would obtain
widely different results. It is not obvious then that when one
uses these potentials in computing the transition matrix elements
that involve low external energy/momentum, one is not getting
different answers depending upon which short-distance properties
are involved in the potentials. This is the difficulty that the
SNPA cannot control by itself.

The MEEFT resolves, albeit approximately, this difficulty as
follows. We start with the wave functions for the initial and
final nuclear states computed in many-body techniques~\cite{SNPA}
using a potential that is fit to a large set of scattering and
transition data. Such a potential typically possesses the
long-range tail given by the pion exchange constrained by chiral
symmetry -- i.e., consistent with low-order chiral expansion --
and the short-range parts determined by the fit to experiments.
While the long-range part is unique, the short-range part may not
be since a variety of short-distance components can give the same
physical low-energy observables. The wave functions obtained with
such different potentials would possess different short-distance
properties, so the matrix element of an arbitrary operator
computed to a finite order will differ depending upon how
sensitive the operator is to the short-distance physics. What is
needed to get the correct matrix element that is independent of
short-distance ambiguities is the additional terms in the operator
that compensate the short distance behavior in the wave functions.
The way to do this is as follows. Since the current appears only
once in the matrix element, one can focus on the irreducible
graphs that in chiral perturbation theory contribute to a finite
order $N$. In calculating the loop graphs, one puts a cutoff in
the loop integrals, say, $\bar{\Lambda}$~\footnote{How to do this
at higher loop order while preserving chiral symmetry is a tricky
matter when one uses the cutoff procedure. To avoid this problem,
one might use the dimensional regularization suitably generalized
for the problem at hand, e.g., power divergence subtraction.} and
integrate out above $\bar{\Lambda}$. What is integrated out gives
a set of contact terms and make up the counter terms in the
renormalization procedure. Those counter terms are mostly unknown
unless the theory is renormalizable or ``defined" by a fit to
experiments. In any case, they will depend on $\bar{\Lambda}$.
Suppose now that the matrix element of the operator so computed to
the order $N$ with the wave functions described above is fit to
experiments, thereby fixing the counter term coefficients for
given $\bar{\Lambda}$. This may or may not be possible depending
upon how many counter terms there are and how many experimental
data are available. Suppose that we can determine $all$ the
counter terms in this way. Then one can use the same set of
operators to compute transitions in other systems, that is to say,
make predictions. Since the long-range parts are unambiguously
fixed by chiral symmetry and the zero-range interactions are
$universal$ in that they should not differ from an $A$-body system
to $A\pm 1$-body system, if one has the operator for the $A$-body
system in this way, we can compute the corresponding transition
matrix elements for the $A\pm 1$-body systems without any unknown
parameters. If the procedure is correct, then the result should
not depend sensitively on the cutoff chosen as long as it is
within a reasonable range dictated by the degrees of freedom that
intervene.

The price to pay in the MEEFT is that the strict power counting is
sacrificed at the order $(N+1)$ since the potential is most likely
to contain not only the terms that are consistent with chiral
symmetry up to order $N$ but also terms which go beyond. The
counting error that shows up at $(N+1)$-st order will then be in
the regime of short-distance physics and the role of the MEEFT
regularization is to largely, if not completely, compensate for
this error.

What happens in the case at hand is very much like the situation
with $V_{lowk}$ discussed by Bogner et al~\cite{kuo} that figures
in the half on-shell T matrix for low-momentum effective nuclear
interactions: There it is found also to be remarkably independent
of the short-distance uncertainty.

In the case of the processes (\ref{pp}) and (\ref{hep}), it is
possible to reduce the only two unknown two-body contact terms
present in the theory into one combination in the Gamow-Teller
operator effective for the two-nucleon, three-nucleon and
four-nucleon processes in question (three-body and higher-body
terms are suppressed compared to the terms calculated). The
resulting single unknown constant can then be fixed once and for
all by the accurately measured triton beta decay for $given$
$\bar{\Lambda}$'s. This allows one to predict, totally
parameter-free, the matrix elements for the $pp$ and $hep$
processes, i.e., the results (\ref{Spp}) and (\ref{Shep}). The
constant depends on $\bar{\Lambda}$ but the total Gamow-Teller
matrix element is not. I find this result truly remarkable. It
will be of help in the future work on solar neutrinos.
\section{EFT for dense matter}
Let me turn now to dense hadronic matter. We will consider dense
matter, ranging from nuclear matter all the way to that relevant
to chiral restoration. For this purpose, we will need what I would
call ``double decimation" strategy which consists of first going
to nuclear matter density and then to higher density near the
chiral phase transition. Here the crucial role is played by the
notion of ``vector manifestation of chiral symmetry" introduced by
Harada and Yamawaki~\cite{HY:VM,HY:PR}. The significance of this
notion is best appreciated if one recognizes that it in essence
provides the mechanism by which the mass of the nucleons -- which
makes up 99\% of the mass of ordinary matter~\cite{wilczek} -- can
be made to ``disappear."
\subsection{Vector manifestation}
In going to nuclear matter and beyond, we must keep the
vector-meson degrees of freedom $explicit$ because of the vector
manifestation (VM) mentioned above. To understand the VM, we
consider the HLS Lagrangian~\cite{bandoetal} in which the pion and
vector mesons are effective degrees of freedom. For the moment, we
ignore fermions and heavier excitations as e.g., $a_1$, glueballs
etc. To make the discussion the most transparent, we consider
three massless flavors. Masses and symmetry breaking can be
introduced with attendant complications. The relevant fields are
the (L,R)-handed chiral fields $\xi_{L,R}=e^{i\sigma/F_\sigma}
e^{\mp i\pi/F_\pi}$ where $\pi$ is the pseudoscalar Goldstone
bosons and $\sigma$ the Goldstone scalar absorbed into the HLS
vector field $\rho_\mu$ coupled gauge invariantly with the gauge
coupling constant $g$. If one matches this theory to QCD at a
scale ${\Lambda_M}$ below the mass of the heavy mesons that are
integrated out but above the vector ($\rho$) meson mass, it comes
out -- when the quark condensate $\la \bar{q}q\ra$ vanishes as in
the case of chiral restoration in the chiral limit -- that
 \be
g(\bar{\Lambda})\rightarrow 0,\ \ a(\bar{\Lambda})\equiv
F_\sigma/F_\pi \rightarrow 1.
 \ee
Now the renormalization group analysis shows that $g=0$ and $a=1$
is the fixed point of the HLS theory and hence at the chiral
transition, one approaches what is called the ``vector
manifestation" fixed point. The important point to note here is
that {\it this fixed point is approached regardless of whether the
chiral restoration is driven by temperature $T$ or density $n$ or
a large number of flavors}. At the VM, the vector meson mass must
go to zero in proportion to $g$, the transverse vectors decouple
and the longitudinal components of the vectors join in a
degenerate multiplet with the pions~\footnote{One might wonder
whether the decoupled transverse vectors require chiral partners
to be consistent with the Wigner mode of chiral symmetry. If so,
then, there would be a difficulty as there are no massless axial
vectors that belong to the same multiplet. The answer is that
there is no need for them. The transverse vector fields must
transform invariantly under chiral transformation. How this can
happen is suggested in \cite{MR:Taiwan}. I am grateful for
discussions with M. Harada, T. Kugo and K. Yamawaki on the status
of the transverse vectors.}. Specifically
 \be
m_\rho^*/m_\rho \approx g^*/g \approx
\la\bar{q}q\ra^*/\la\bar{q}q\ra \rightarrow 0
 \ee
as the transition point $n=n_c$ is reached~\cite{HKR}~\footnote{We
will specialize in density. Qualitatively the same argument holds
also in temperature but the details can be different away from the
chiral restoration point.}. This result can be understood as
follows. Near the critical point the ``intrinsic term" $\sim g^*
F_\pi^*$ in the vector mass formula drops to zero faster than the
dense loop term that goes as $\sim g^* H(n)$ where $H$ is a slowly
(i.e., logarithmically) varying function of density. So the dense
loop term controls the scaling. Now it seems to be a reasonable
thing to do to assume that near the VM fixed point, we have the
scaling
 \be m_\rho^*/m_\rho \approx g^*/g\approx
\la\bar{q}q\ra^*/\la\bar{q}q\ra.
 \ee
Our conjecture~\cite{BR-DD} is that this holds from nuclear matter
density to the chiral restoration point.
\subsection{Double-decimation conjecture}
Let us now turn to the low-density regime, that is, density below
nuclear matter density. At near zero density, one can apply chiral
perturbation theory with a zero-density chiral Lagrangian. Using
the HLS Lagrangian matched to QCD at a scale $\Lambda_M\sim
\Lambda_\chi$, one can establish that
 \be
m_\rho^*/m_\rho \approx f_\pi^*/f_\pi \approx
\sqrt{\la\bar{q}q\ra^*/\la\bar{q}q\ra}.\label{dd1}
 \ee
This result follows from an in-medium GMOR relation for the pion
with the assumption that at low density the pion mass does not
scale (as indicated experimentally~\cite{yamazaki}), that the
vector meson mass is dominantly given by the ``intrinsic term"
$\sqrt{a} F_\pi g$ with small loop corrections that can be ignored
and that the gauge coupling constant does not get modified at low
density (as indicated by chiral models and also empirically). We
will now simply assume that this relation holds from zero density
to nuclear matter density. An important ingredient to interject
here is that normal nuclear matter is a many-body fixed point,
known as Fermi-liquid fixed point, so we are essentially
summarizing the phase structure by two fixed points, namely, the
Fermi-liquid fixed point and the vector-manifestation fixed point.
This suggests a double-decimation approximation. It may be too
simplistic to approximate the complex phase structure by only two
decimations but we find that we can go a long distance with this
approximation.
\section{Predictions}
A variety of evidences that lend support to the scaling behavior
given above are discussed in \cite{BR:PR01}. Here I would like to
discuss a few recent developments.
\subsection{$Parametric$ density dependence}
The one feature that distinguishes the HLS/VM theory from other
EFTs is the $parametric$ dependence on the background of the
``vacuum" -- density and/or temperature -- which controls the
fixed point structure of the VM. At low density, this dependence
is relatively weak but in precision experiments, it should be
visible.

Consider a chiral Lagrangian in which only the nucleon and pion
fields are kept explicit with the vectors and other heavy hadron
degrees of freedom integrated out (say, from the HLS/VM
Lagrangian). The relevant parameters of the Lagrangian are the
``bare" nucleon mass $M(\Lambda_M,n)$, the ``bare" pion mass
$m_\pi (\Lambda_M,n)$, the ``bare" pion decay constant $F_\pi
(\Lambda_M,n)$, the ``bare" axial-vector coupling $g_A
(\Lambda_M,n)$ and so on. The structure of this Lagrangian is the
same as the familiar one apart from the $intrinsic$ dependence of
the parameters on $n$. In the usual approach, the scale
$\Lambda_M$ is fixed at the chiral scale and the dependence on $n$
is absent. Suppose that one does a chiral perturbation theory with
the HLS/VM Lagrangian. The power counting will be the same as in
the conventional approach. Now if the density involved in the
system is low enough, say, no greater than nuclear matter density,
then one could work to leading order in chiral expansion. Suppose
that one does this to the tree order. To this order, the
parameters of the Lagrangian can be identified with physical
quantities. For instance, the bare pion decay constant $F_\pi$ can
be identified with the physical constant $f_\pi$ etc. Now in the
framework at hand, the only dependence in the constants on density
will then be the $intrinsic$ one determined by the matching to QCD
living in the background of density $n$.

If we apply the above argument to the recent measurement of deeply
bound pionic atom systems~\cite{yamazaki}, we will find that the
measurement gives information on the ratio $f_\pi^*/f_\pi$ at a
density $n\lesssim n_0$. We have from (\ref{dd1})
 \be
\Phi (n)\equiv f_\pi^*/f_\pi \approx
\sqrt{\la\bar{q}q\ra^*/\la\bar{q}q\ra} (\approx
m_\rho^*/m_\rho).\label{dd2}
 \ee
The scaling quantity $\Phi$ has been obtained from nuclear
gyromagnetic ratio in \cite{FR,BR:PR01}. At nuclear matter
density, it comes out to be $\Phi (n_0)=0.78$ with a small error
bar. Thus it is predicted that
 \be
(f_\pi^*/f_\pi)^2 (n_0)\approx 0.61.
 \ee
This agrees with the value extracted from the pionic atom data of
\cite{yamazaki}, $(f_\pi^*/f_\pi)^2 (n_0)=0.65\pm 0.05$.

It is perhaps important to stress that this ``agreement" can be
taken as an evidence neither for the validity of the tree
approximation nor for the signal for ``partial chiral
restoration." If one were to go to higher orders in chiral
expansion, the $parametric$ pion decay constant cannot be directly
identified with the physical pion decay constant since the latter
should contain two important corrections, i.e., quantum
corrections governed by the renormalization group equation as the
scale is lowered from $\Lambda_M$ to the physical scale and dense
loop corrections generated by the flow. At the chiral restoration,
it is this latter that signals the phase transition: The
parametric pion decay constant with the scale fixed at the
matching scale does not go to zero even at the chiral restoration
point~\cite{HY:PR}. Thus when one does higher-order chiral
perturbation calculation of the same quantity, one has to be
careful which quantity one is dealing with.
\subsection{Effective degrees of freedom at the chiral phase
transition} Let me go near the chiral transition point and ask
what new physics is introduced by the VM. I will do this for high
temperature. There are lots of interesting phenomena triggered by
high density in the present framework but this will be a topic for
next publication.

Let me start by recalling the standard scenario of the chiral
phase transition driven by temperature. The standard scenario
assumes that in the heat bath, there are no other low-energy
degrees of freedom than pions to consider, so all other mesons and
all baryons can be integrated out with their imprints lodged in
the coefficients of higher-dimension operators in the Lagrangian.
In making this assumption, one is disregarding the possible
relevance of the VM fixed point. Consider now the vector
susceptibility (VSUS) and axial-vector susceptibility (ASUS)
defined in terms of Euclidean QCD current correlators as
 \be
\delta^{ab}\chi_V&=& \int^{1/T}_0 d\tau\int d^3\vec{x}\la V_0^a
(\tau, \vec{x}) V_0^b (0,\vec{0})\ra_\beta,\\
\delta^{ab}\chi_A&=& \int^{1/T}_0 d\tau\int d^3\vec{x}\la A_0^a
(\tau, \vec{x}) A_0^b (0,\vec{0})\ra_\beta
 \ee
where $\la \ra_\beta$ denotes thermal average and
 \be
V_0^a\equiv \bar{\psi}\gamma^0\frac{\tau^a}{2}\psi, \ \
A_0^a\equiv \bar{\psi}\gamma^0\gamma^5\frac{\tau^a}{2}\psi
 \ee
with the quark field $\psi$ and the $\tau^a$ Pauli matrix the
generator of the flavor $SU(2)$. These SUS's can be readily
computed via chiral Lagrangians defined in a heat bath with the
sole assumption that the effective Lagrangian $\L_{eff}$ can be
written in terms of local fields and their derivatives only -- an
assumption which is standard in EFTs. Indeed, Son and
Stephanov~\cite{son} computed the ASUS, $\chi_A$, and found at the
critical temperature and in zero density that \be
\chi_A=-\frac{\del^2}{\del\mu_A^2}\L_{eff}|_{\mu_A=0}={f_\pi^t}^2
 \ee
where $f_\pi^t$ is the time component of the pion decay constant.
The principal point to note here is that {\it as long as the
effective action is given by local terms,  this is the entire
story}: There are no other contributions to the ASUS than the
temporal component of the pion decay constant. Now at the chiral
phase transition, the vector SUS and the axial-vector SUS must be
equal to each other, $\chi_V=\chi_A$ and the lattice data on
$\chi_V$~\cite{gottlieb} clearly indicates that
 \be
\chi_V|_{T=T_c}\neq 0,
 \ee
which leads to the conclusion~\cite{son} that
 \be
f_\pi^t|_{T=T_c}\neq 0.
 \ee
On the other hand, it is expected and verified by lattice
simulations that the space component of the pion decay constant
$f_\pi^s$ should vanish at $T=T_c$. One therefore arrives at
 \be
v_\pi^2\sim f_\pi^s/f_\pi^t\rightarrow 0, \ \ T\rightarrow T_c
 \ee
where $v_\pi$ is the pion velocity. This is the main conclusion of
the pion-only theory.

This elegant result is unfortunately marred by the caveat that the
same formalism applied to the VSUS does not work. In fact, one can
see easily from the effective Lagrangian that the same theory
predicts $\chi_V=0$ for $all$ temperature which of course cannot
be correct and is ruled out by the lattice result. Son and
Stephanov attribute this failure to the possibility that a
diffusive mode in hydrodynamic language, which is not described by
the chiral Lagrangian, contributes to the VSUS. In the standard
picture, this is not surprising since there are no other degrees
of freedom than the pions and it must be nonlocal effects
involving pions that must intervene. This situation is entirely
different when the vectors are present as relevant degrees of
freedom as in the VM.

In the VM scenario, the vector mesons with a dropping mass must
enter near the phase transition. In addition, one can conceive of
quasiquarks with a similarly dropping mass also contributing. It
is rather straightforward to compute the VSUS and ASUS at one-loop
order. The results obtained by Harada, Kim, Rho and
Sasaki~\cite{HKRS} are
 \be
f_\pi^t|_{T=T_c}=f_\pi^s|_{T=T_c}=0, \ \
v_\pi|_{T=T_c}=1\label{main1}
  \ee
and
 \be
\chi_A|_{T=T_c}=\chi_V|_{T=T_c}= 2N_f \left[\frac{N_f}{12} T_c^2
+\frac{N_c}{6} T_c^2\right] \ .\label{main2}
 \ee
These results are not difficult to understand. Both the time part
and space part of the pion decay constant go to zero with the
ratio going to 1. Thus the pion velocity approaches the speed of
light, which is completely opposite to the Son-Stephanov result of
vanishing velocity. The vector mesons and the quasiquarks
contribute to the vector and axial-vector SUS's, the latter bigger
than the former by a factor of 2. It is easy to see how the two
SUS's come out equal here. Both are given by quasiparticles, the
vector mesons and constituent quarks. There is no need to invoke
diffusive modes.

It is interesting that both pion decay constants go to zero
simultaneously giving the pion velocity equal to the light
velocity. It appears to restore Lorentz symmetry broken by the
medium. It is however counterintuitive: one could think of the
pion as an analog to the sound in condensed matter, the velocity
of which is known to go to zero on critical surface. The VM result
looks elegant and simple but is puzzling at the moment.

There is also a caveat here. The result obtained for the pion
velocity is in one-loop approximation. One might wonder what
happens at higher-loop orders. We expect the $f_\pi^s$ to remain
zero at the critical point to all orders. This is because it is
the order parameter for the chiral restoration. The question is
what happens to the time part, $f_\pi^t$, at higher orders. It is
possible that the VM fixed point protects it so that it remains
zero but this has not been proven. If however there is an
additional contribution at higher order that does not vanish, then
we will go over to the Son-Stephanov result of a vanishing pion
velocity. More work needs to be done to clarify this issue.

On the other hand, if (\ref{main1}) does turn out to be correct,
then the implication will be that the phase structure at the
chiral transition is entirely different in the HLS/VM from that of
the standard picture. In fact it indicates that there can be a
much richer structure in the phase diagram than the standard one.
In particular, it indicates that a certain number of degrees of
freedom hitherto ignored must be taken into account and the
transition will be a lot smoother than imagined in the standard
scenario since the number of degrees of freedom below the critical
point will be considerably increased. I expect that future lattice
and experimental developments will validate or invalidate this
beautiful VM scenario.

\end{document}